\documentclass[a4paper,floatsintext, jou, 10pt, natbib]{apa6}

\usepackage[english]{babel}
\usepackage[utf8x]{inputenc}
\usepackage{amsmath}
\usepackage{graphicx}
\usepackage{float}
\usepackage{setspace}
\usepackage{amsmath}
\usepackage{color}

\newcommand{\hamit}[1]{\textcolor{black}{#1}}
\providecommand{\keywords}[1]{\textbf{\textit{Index terms---}} #1}

\title{Predictive Event Segmentation and Representation with Neural Networks: A Self-Supervised Model Assessed by Psychological Experiments}

\shorttitle{Event Segmentation}
\author{Hamit Basgol\textsuperscript{1}, 
        Inci Ayhan\textsuperscript{2}, 
        Emre Ugur\textsuperscript{3}}

\affiliation{Department of Computer Science\textsuperscript{1} at the University of Tübingen, Germany\break
            Department of Psychology\textsuperscript{2},\break
            Department of Computer Engineering\textsuperscript{3}
            at Boğaziçi University, Turkey}

\abstract{People segment complex, ever-changing and continuous experience into basic, stable and discrete spatio-temporal experience units, called events. Event segmentation literature investigates the mechanisms that allow people to extract these units from the continuous experience. Aiming to shed light on event segmentation ability, event segmentation theory points out that people predict ongoing activities and observe prediction error signals in order to find event boundaries that keep events apart. In this study, we investigated the mechanism giving rise to this ability by a computational model and accompanying psychological experiments. Inspired from the principles of event segmentation theory and predictive processing, we introduced a semi-mechanistic model of event segmentation, learning, and representation. This model consists of feed-forward neural networks that predict the sensory signal in the next time-step in order to represent different events, and a cognitive model that regulates these neural networks on the basis of their prediction errors. In order to verify the ability of our model in segmenting experience into spatio-temporal units, learning them during passive observation, and representing them in its internal representational space, we prepared a video that depicts natural human behaviors represented by point-light displays. We compared event segmentation behaviors of human participants and our model with this video in two hierarchical event segmentation levels. By using point-biserial correlation technique, we demonstrated that event segmentation decisions of our model correlated with the responses of participants. Moreover, by approximating internal representation space of participants by a similarity-based technique, we showed that our model formed a similar internal representation space with those of participants. Our results suggests that our model that tracks the prediction error signals can produce human-like event segmentation decisions and event representations. Finally, we discussed our contribution to the literature of event cognition and our understanding of how event segmentation is implemented in the brain.}


\begin{document}
\maketitle

\footnotetext[1]{Hamit Basgol is a PhD Student in the Department of Computer Science at the University of Tübingen, Tübingen, Germany (e-mail: hamitbasgol@gmail.com). The study is conducted when Hamit Basgol is a master's student in Cognitive Science Department at Bogazici University.}
\footnotetext[2]{Inci Ayhan is with Department of Psychology in Bogazici University, Istanbul, Turkey (e-mail: inci.ayhan@boun.edu.tr)}
\footnotetext[3]{Emre Ugur is with Department of Computer Engineering in Bogazici University, Istanbul, Turkey (e-mail: emre.ugur@boun.edu.tr)}

\section{Introduction}

Humans segment continuous information stream into event units to show robust, adaptive, and intelligent behavior, which is called event segmentation \citep{Zacks2007EventPerception, Zacks2007EventSegmentation, Zacks2020EventPerception, Richmond2017Constructing}. In the recent years, a growing number of computational models were proposed to capture how humans segment events in order to utilize their continuous experience \citep{Franklin2020Structured, Gumbsch2016Learning, Gumbsch2017ComputationalModelDyn, Reynolds2007Computational, Metcalf2017Modelling}. Despite their essential contribution to the understanding of the event segmentation ability, these models have demonstrated certain limitations. Namely they were not capable of segmenting events in varying lengths \citep{Reynolds2007Computational, Metcalf2017Modelling}, they used datasets that involved abrupt transitions between naturalistic action sequences \citep{Reynolds2007Computational, Metcalf2017Modelling} and they included robotic models that did not aim for capturing human event segmentation decisions \citep{Gumbsch2016Learning, Gumbsch2017ComputationalModelDyn}. In fact, to the best of our knowledge, there has been only one study, where authors compared the performance of their model to the human event segmentation decisions \citep{Franklin2020Structured}.

In this study, we aim at addressing these limitations by a novel computational model, which we built upon three main elements: (1) the event segmentation theory \citep{Zacks2007EventPerception}, (2) the predictive processing of \cite{Wiese2017Vanilla, Clark2013Whatever}, and (3) the robotic model of \cite{Gumbsch2017ComputationalModelDyn}, the contributions of which will be highlighted throughout the paper. In this study, we showed that a self-supervised and semi-mechanistic model monitoring prediction error signals could produce multimodal event segments in varying lengths and store the knowledge of events in activations of neural networks. Moreover, we compared the segmentation and representation results of our model with those of humans to reveal their similarities and differences. We believe that our model presents a fruitful approach to modeling event segmentation and integrating event knowledge into a wide range of perceptual and cognitive processes.

The introduction of the paper is organized as follows: Firstly, we introduce the event segmentation theory and the importance of the prediction error signals for the event segmentation. Secondly, we review the computational models of event segmentation and identify their limitations. Finally, we explain the methodology of our current study, its results and conclusions.

\subsection{Event segmentation theory and prediction error}

Early studies of event segmentation were conducted by \cite{Newtson1973Attribution} using a unitization paradigm, where participants were asked to watch a movie and segment it into meaningful units. The results of Newston's study demonstrated that a substantial agreement across participants on the segmentation locations, which happened to be persistent in time. Subsequent research verified these findings and opened up the possibilities of investigating the role of events in human cognition \citep{Zacks2007EventSegmentation, Zacks2020EventPerception}. The locations at which participants segment a continuous information stream (e.g., a movie) are termed as event boundaries, which are the positions in time that show perceptual changes in spatial locations, movements, relative distances between agents, or goals \citep{Kurby2008Segmentation, Hard2011Shape, Zacks2007EventPerception, Zacks2020EventPerception, Cutting2012Perceiving, Hard2006Making, huff2014changes, Newtson1977Obj, Zacks2010Brain, cutting2014event}. 

Events are known to be hierarchically structured \citep{Zacks2020EventPerception}. People can detect smallest (fine-grained) and largest events (coarse-grained) \citep{Newtson1973Attribution, Hard2006Making, Hard2011Shape, Zacks2007EventSegmentation, Zacks2020EventPerception, Zacks2001HumanBrain}, when they are instructed to do so. Research with functional magnetic resonance neuroimaging (fMRI) suggests that hierarchical segmentation is an automatic process \citep{Zacks2001HumanBrain, Speer2007HumanBrain} such that while observing a movie or reading a story, the brain selectively responds to the fine- and coarse-grained event boundaries.  \cite{Hard2011Shape}, for example, demonstrated that changes at event boundaries are more numerous than other parts of an activity; moreover, they particularly peak at coarse-grained boundaries. The strong relationship between both types of change, namely the sensory (fine-grained) and the conceptual (coarse-grained) change, suggests that events are segmented based on the perceptual cycle formed by the bottom-up processing of sensory features and the top-down processing of conceptual knowledge \citep{Neisser1976Cognition, Zacks2007EventPerception, Zacks2020EventPerception}.

A computational model or a theory of event segmentation should explain at least two basic properties of event segmentation. The first one is how locations of event boundaries are detected and the second one is how event segmentation operation is conducted in different hierarchies. Event segmentation theory (EST) proposes an account for both of these properties. According to the EST, people constantly make perceptual predictions by event models in the working memory \citep{Zacks2007EventPerception, Reynolds2007Computational}. The event boundary is formed when the current event model cannot capture the current situation, in other words, when the corresponding prediction error signal follows a transient increase. In such situations, the system triggers another event model to predict the following sensory input. Thus, a strategy that is based on monitoring the prediction error signals, might correspond to the basic mechanism behind the event segmentation ability. Indeed, the EST and the role of prediction error signals in the event segmentation were supported by many studies \citep{Zacks2011Prediction, Eisenberg2018DynamicPrediction, Hard2011Shape, Gumbsch2016Learning, Gumbsch2017ComputationalModelDyn, Franklin2020Structured, Reynolds2007Computational, Stawarczyk2021Event}, despite exceptions \citep{Shin2021Structuring, OReilly2013Making}. Along with its focus on the prediction error signals for event boundary detection, EST also suggests that people might make predictions by events in multiple timescales simultaneously and sensitivity differences between events to incoming prediction error signals might determine their lengths or positions in the hierarchy. For example, an event model might be sensitive to minor prediction errors compared to another \citep{Zacks2007EventSegmentation}, and this sensitivity difference might make the former shorter than the latter.

Due to this mechanism, computational models in the literature have been mostly inspired from the EST \citep{Franklin2020Structured, Gumbsch2016Learning, Gumbsch2017ComputationalModelDyn, Reynolds2007Computational, Metcalf2017Modelling}. All these models, on the other hand, come up with certain limitations, which will be the topic of the next sub-section.

\subsection{Computational models of event segmentation}

Several important computational models have been proposed in the literature with different limitations \citep{Franklin2020Structured, Gumbsch2016Learning, Gumbsch2017ComputationalModelDyn, Reynolds2007Computational}. For example, \cite{Reynolds2007Computational} utilized a set of sequence models for the segmentation of human behaviors. Despite the success of the model in detecting event boundaries, hierarchical segmentation of events in varying granularities was not addressed. At the same time, behavioral sequences that were used for training the model involved abrupt and unnatural transitions. \cite{Metcalf2017Modelling} enhanced this model by a reinforcement learning agent. Although these two models suggest that monitoring prediction error signals is an effective strategy for the event segmentation, they did not address hierarchical segmentation of events.

\cite{Gumbsch2016Learning, Gumbsch2017ComputationalModelDyn} developed a robotic model that chunks sensory-motor information flow into parts. The model represents events by linear models, which encode different sensory dimensions and predict sensory signal in the next time-step. The linear models are regulated by a cognitive model at a higher level based on the prediction errors of the lower-level linear models. From this perspective, whereas the cognitive model resembles to the mechanism proposed by the EST, linear models correspond to the working memory representations. As has been addressed in \cite{Gumbsch2019Autonomous}, however, since linear models encoding sensory dimensions are disconnected from one another, Gumbsch et al.'s model is not capable of discovering multi-modal associations between sensory modalities in a particular event structure. Besides this limitation, these models are robotic models that assume the involvement of an active agent. However, the event segmentation literature is based largely on the unitization paradigm in which participants observe events passively and press a button to separate them from one another \citep{Newtson1973Attribution, Newtson1976Perceptual, Newtson1977Obj, Zacks2007EventPerception, Zacks2020EventPerception, Hard2003Segmenting, Hard2006Making, Zacks2001HumanBrain}.

Lastly, \cite{Franklin2020Structured} developed an inclusive model of event cognition, which considers various domains such as event memorization, segmentation, retrieval, and inference. To the best of our knowledge, this is the only study that used naturalistic videos and considered the human-level event segmentation performance for the model validation, even though the received correlation between the performances of the model and ground-truth data is open to improvement.

Overall, the computational models of event segmentation suggest that the EST presents a plausible mechanism for the event segmentation task. For addressing the missing points in the literature and testing the suggested mechanism by the EST for more than one granularity level, we developed a novel computational model for event segmentation. In addition to event segmentation capability, our model could form event representations.

\subsection{Event representations}

Representations are mental objects with semantic properties \citep{Pitt2020MentalRepresentation}. To express the strength of the relationships, a representational space can be formed by taking the pairwise distance between all representations \citep{Shepard1979Additive, Shepard1980Multidimensional, Shepard1987Toward}. This two-way relationship makes the similarity a valuable metric to reveal how the system organizes knowledge as representations form the basis of categorization and generalization. 
One aim of the artificial intelligence is to learn valuable and representative information from the data \citep{Bengio2014Representation}. Multi-layer perceptrons (i.e., deep neural networks) can learn distributed and semantically meaningful representations \citep{Bengio2014Representation, Urban2019DeepLearning}. The similarity between representations (i.e., semantic relationships between represented entities) of a deep learning model can be found by the Euclidean distance or cosine similarity. For example, the semantic relationship between words and sentences \citep{Mikolov2013Distributed, Rogers2005Parallel}, objects \citep{Deselaers2011Visual}, scenes \citep{Eslami2018Neural}, and episodes \citep{Rothfuss2018DeepEpisodic} can be captured with the help of representations learned by a deep learning system. Since representations give researchers a gist about how humans organize knowledge, generalize between instances, and make analogical transfers \citep{Blough2001Perception, Nosofsky1992Similarity, Shepard1987Toward, Shepard1980Multidimensional, Tversky1977Features}, they have a fundamental place in cognitive science. As could be expected, researchers exploited human similarity judgments to achieve human mental representations \citep{Shepard1980Multidimensional, Shepard1987Toward, Shepard1979Additive}. The role of representations and similarity judgments in artificial intelligence and cognitive science suggest that they might provide a basis for comparing people and machines. In fact, recent research provides excellent examples of this comparison \citep{Peterson2018Evaluating, Hebart2020Revealing}. 

Event representation literature is very rich and represents a diverse set of studies \citep{Blom2020Predictions, Schutz-Bosbach2007ProspectiveCoding, Day2008Representation, Wang2017Predicting, Sheldon2018CognitiveTools, Fivush1992Structure, Kominsky2020Causality}. In the context of computational modeling, recent studies use \citep{Shen2020Study} and learn \citep{Dias2019Learning} event representations. In contrast, despite the interest received by event representations, event similarity judgments is a concealed area under the action similarity judgments \citep{Tarhan2020Semantic, Tarhan2018HighLevel}. In our work, utilizing this possibility, we compare the event representations of our computational model and participants by exploiting event similarity judgments.

\subsection{Our contribution}

In this study, inspired from the EST \citep{Zacks2007EventPerception}, the predictive processing \citep{Wiese2017Vanilla, Clark2013Whatever}, and the Gumbsch's robotic model \citep{Gumbsch2016Learning, Gumbsch2017ComputationalModelDyn}, we developed a novel computational model for event segmentation. Our model consists of multi-layer perceptrons (i.e., event models) that are managed by a cognitive mechanism and consequently, determining the event boundaries. \hamit{As our contribution to the literature, (1) our model is capable of learning to represent and predict multi-modal event segments with sensory associations in passive observation unlike the models developed by \citep{Gumbsch2016Learning, Gumbsch2017ComputationalModelDyn} which segment unimodal events based on actions in a simulation environment. (2) With the help of a parameter, changing sensitivies of event models to prediction error signals, our model can also segment events in varying granularities, which was not addressed by \cite{Reynolds2007Computational} and \cite{Metcalf2017Modelling}. (3) Moreover, segmentation and representation capabilities of our model were tested by ground-truth data received from psychological experiments.}

A multi-layer perceptron is a plain deep neural network which consists of an input layer, an intermediate (hidden) layer or layers, and an output layer. The network learns the relationship between inputs and outputs by updating weights in each iteration. Thanks to the hidden units, multi-layer perceptrons can classify complex patterns \citep{lippmann1989pattern}, approximating non-linear functions \citep{hornik1989multilayer}. Moreover, the knowledge developed throughout the training is stored in weights and what is learned by the model can be explored by analyzing the representations of the network \citep{Peterson2018Evaluating, Hebart2020Revealing, fleming2019learning}. The use of deep neural networks in cognitive science and artificial intelligence has a long history and had an important role in the emergence of the connectionist framework \citep{rumelhart1986learning}. The effect of connectionist framework, in other words deep learning models, on cognitive sciences still persists and leads to revolutionary results in a wide variety of domains such as perception \citep{krizhevsky2012imagenet, russakovsky2015imagenet, he2016deep, spoerer2017recurrent, fleming2019learning}, linguistics \citep{radford2017learning, wu2016google, floridi2020gpt}, developmental psychology \citep{orhan2020self}, and cognitive neuroscience \citep{yamins2016using, khaligh2014deep, tripp2017similarities}.

We used multi-layer perceptrons as a member of deep neural networks to represent events. A deep neural network model can be trained in several ways: supervised, unsupervised and self-supervised way. In the supervised learning, models receive outputs (e.g., categories for object identification) of inputs (e.g., images) from huge labelled datasets \citep{krizhevsky2012imagenet, russakovsky2015imagenet}. Despite its success in a range of domains such as object identification \citep{krizhevsky2012imagenet, russakovsky2015imagenet}, supervised learning is criticized being inconsistent with how humans actually learn. Humans learn new concepts and abilities with little supervision without the requirement of hand-crafted labels \citep{vinyals2016matching}. The dependency of supervised learning on labels leads to researchers investigate other learning possibilities such as unsupervised and self-supervised learning that do not require explicit labels. In the unsupervised learning, the network learns how to represent data efficiently by not relying on labels, but rather by capturing the high-order statistics of the dataset \citep{fleming2019learning}. On the other hand, in the self-supervised learning, labels are substituted by the information in the input data so much so that rather than mapping inputs to the hand-crafted labels, the network learns to predict selected parts of the data (e.g., predicting the next sequence of a video or a certain part of an image) and generates representations by this way \citep{liu2021self}.

In the context of event segmentation, using a supervised model receiving event boundary locations from a supervisor or instructor would not be natural because infants mostly learn new abilities without supervision. Therefore, our model is self-supervised as it learns to make prediction in a segment and detects event boundaries from the data without the need of human-crafted labels \citep{liu2021self}.

Despite the contribution of deep neural networks to the progress in a wide range of areas from linguistics to neuroscience \citep{krizhevsky2012imagenet, russakovsky2015imagenet, he2016deep, spoerer2017recurrent, fleming2019learning, radford2017learning, wu2016google, floridi2020gpt, orhan2020self, yamins2016using, khaligh2014deep, tripp2017similarities}, they have an important limitation, namely explainability. Deep neural networks are black-box models whose functioning is not explicit, which is a crucial property especially for the domains where understanding the decisions of networks is critical (e.g., medical decision making) or in scientific practice (e.g., why the network decides this way and what this decision says for the problem in question). Even though it uses multi-layer perceptrons for representing the event segments, our model is not an end-to-end black box model \citep{rudin2019we}; rather, it involves an easily understandable and trackable white-box model that regulates multi-layer perceptrons. From this perspective, our model is a semi-mechanistic model, incorporating the capabilities of white-box and black-box models, giving researchers a chance to benefit from the power of neural networks without fully sacrificing the explainability.

\hamit{
Our proposed model is both self-supervised and semi-mechanistic. By tracking the prediction error signals, (1) the model produces multimodal event segments in varying hierarchies via passive observation with the help of multi-layer perceptrons unlike the models developed by \citep{Gumbsch2016Learning, Gumbsch2017ComputationalModelDyn}, (2) With the help of a parameter, changing sensitivities of event models to prediction error signals, our model can produce event segments in varying granularities, which was not addressed by \citep{Reynolds2007Computational, Metcalf2017Modelling}. (3) Moreover, not only did we study the activations in the layers to have an insight with respect to the functioning, we also compared the performance of our model to that of the human observers in order to assess its capabilities. We received a higher point-biserial correlation score than the existing score in the literature \citep{Franklin2020Structured}. More specifically,}

\begin{itemize}
\item we prepared two videos depicting naturalistic human behaviors that are represented as point-light displays (PLDs) \citep{Johansson1973VisualPerception}. PLDs depict biological movements by several points that are placed on the positions corresponding to joints. From their movements, people can recognize human movements, emotions, actions, and gender \citep{Alaerts2011Action, Troje2008BiologicalMotion}. The ability to perceive PLDs emerges very early in human life \citep{bertenthal1987perception, fox1982perception} and human brain imaging studies \citep{grossman2000brain, krakowski2011neurophysiology, pavlova2004dissociable, michels2009brain, peuskens2005specificity} suggest that biological motion processing are specialized. Since they can provide an easily manipulable stimulus for the psychological experiments and present an opportunity to reduce data dimension and processing time for computational experiments. 

\item By using the dataset prepared from PLDs, we assessed the smallest (fine-grained) and largest (coarse-grained) event segmentation performances of our model. Presenting the same videos to the participants in an online psychological experiment, we collected the behavioral event boundary judgments, to which we then compared the performances of our model by a point-biserial correlation.

\item We proposed a new validation technique inspired from the literature on the concept emergence in deep neural networks \citep{Peterson2018Evaluating} for investigating whether event segments that are produced by our model is meaningful and that neural networks capture human-like event representation space. For approximating event representations of participants, in another online experiment, we asked participants the perceived similarities between different events to turn this into an event representation space. We then compared the internal event representation space of our model with that of the participants. The techniques by which we approximated event representations of our model will be further explained in the method section. 

\end{itemize}

Our results suggested that the proposed model received a considerable point-biserial correlation score ($r_{pb}$) with the event segmentation decisions of participants, namely 0.254 and 0.196 for the fine- and the coarse-grained segmentation, respectively. Moreover, the correlation between the internal representation spaces of the participants and our model resulted in $r$ of 0.435 and 0.614 for fine-grained and coarse-grained segmentation. These results, overall, propose that (1) a model guided by prediction error signals can capture event segmentation decisions of human participants, (2) multi-layer perceptrons can successfully capture the internal representation space, and (3) events can be segmented to one another by a self-supervised predictive model that does not receive event boundary locations as human-crafted labels.

\section{General Method}

In this section, we first illustrate our proposed computational model for the event segmentation and then explain how we prepared the dataset for the psychological and computational experiments.
 
\subsection{The overview of the proposed computational model}

The model we propose is built on top of the computational architecture proposed by \citep{Gumbsch2016Learning, Gumbsch2017ComputationalModelDyn} for the robotic event segmentation. We reformulated the model in question \citep{Gumbsch2016Learning, Gumbsch2017ComputationalModelDyn}. In general (1) we represented event models as multi-layer perceptrons capable of representing non-linear relationships in the data. In this way, our model can now approximate complex sequences of human actions and learn associations between possible modalities. Additionally, (2) we excluded the role of action (i.e., motor modality) in the event formation to extend the model to passive segmentation. 

\begin{figure*}[!ht]
\centering
  \includegraphics[scale = 0.4]{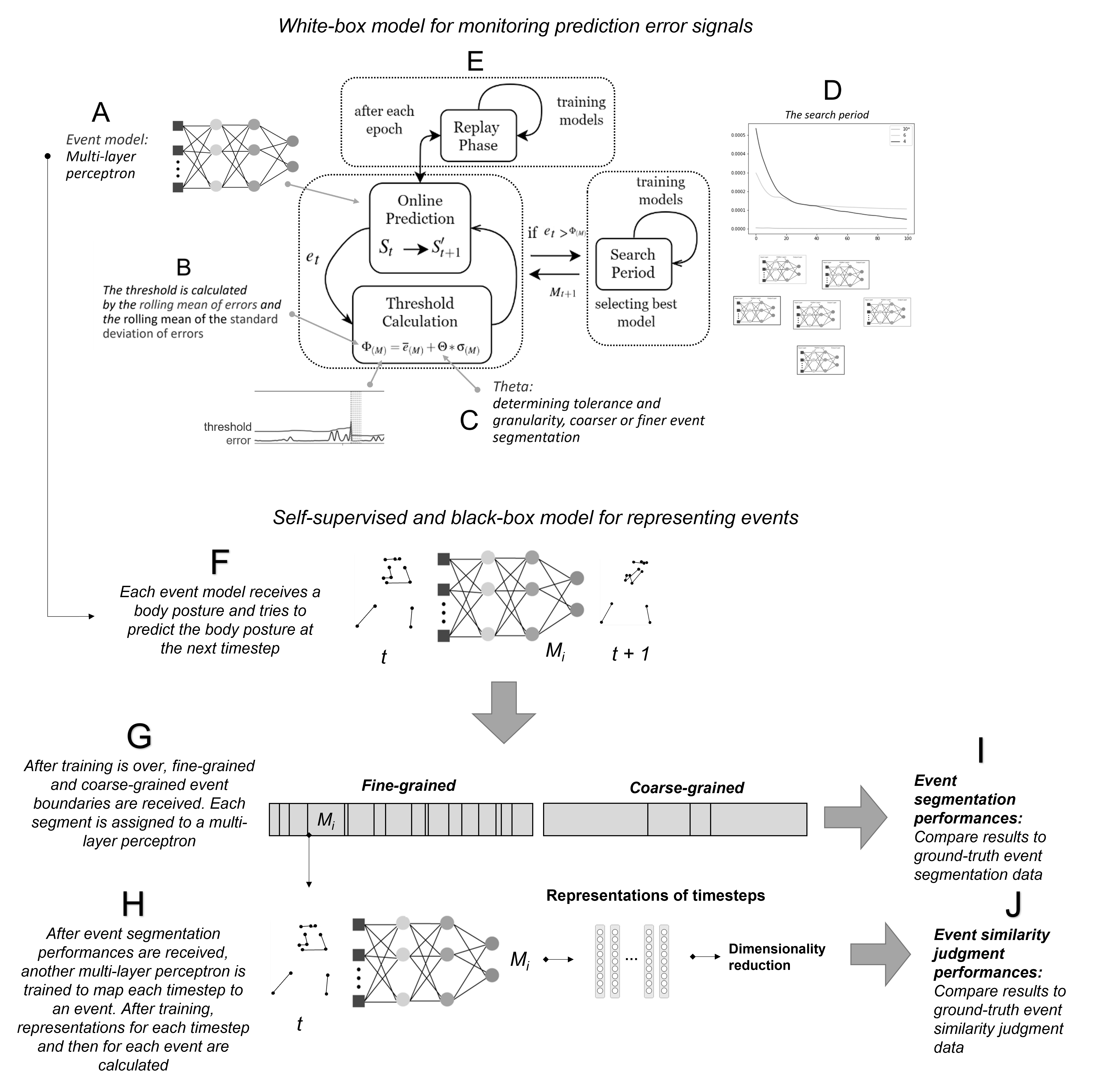}
    \caption{The overview of the study\break(A) In the online prediction phase, the current event model makes a prediction, on the basis of which the current prediction error is calculated. (B) At the same time, for each model, a surprise threshold is computed. If the prediction error exceeds the surprise threshold, the system enters into a search period for spotting the best event model. (C) The parameter $\theta$ determines the tolerance of event models, and therefore, their granularities. (D) In the search period, all event models are trained for the next n timesteps which returns the best event model. (E) After each epoch, the system moves into the replay phase in which all event models are trained for the timesteps for which they are active. (F) Each event model receives a body posture represented by the PLDs and tries to predict the body posture at the next timestep. Event models are multi-layer perceptrons whose functioning is not explicitly tractable. (G) After the training is over, the segmentation results are calculated, and (I) they are compared with human performance obtained from the online psychological experiments. (H) In order to investigate whether segmented events are meaningful and their semantic relationships can be captured by a hierarchical neural network, we received representations of each timestep and reduced their dimensionality. (J) Similarity judgments extracted from these representations are then compared with those of participants.}
\label{fig:computational_model_process}
\end{figure*}

The overview of the proposed model and how the proposed model is tested are given in Figure ~\ref{fig:computational_model_process}. The model represents events as predictive models that predict sensory information at the next timestep and receive a prediction error signal (A, F). Prediction error signals are accumulated and used for calculating a surprise threshold (B), whose tolerance, therefore granularity, is weighted by a parameter called $\theta$ (C). If the surprise threshold is exceeded, the searching period starts. From this period, the best event model is returned (D). After each to be chunked event ends, all activated events during the last training step are trained (E). For subsequent analyses, event segmentation results are calculated (G) and compared with human performance (I). For further investigations, a hierarchical neural network is trained to capture the relationship between events (H), and its results are compared with human performance (J).

\subsubsection{Online prediction}

Our model continuously makes predictions by using an active event model ${M_t}$ at the time point ${t}$ (see Figure ~\ref{fig:computational_model_process}A). $M_{t}$ predicts the next sensory observation by

\[S'_{t + 1} = S_{t} + \Delta S'_{t + 1}\]

Making predictions until receiving the next surprise signal, the active model ${M_t}$ receives sensory inputs, computes and stores prediction error signals, and learns to make predictions via backpropagation. At each prediction ${M_t}$ makes, the system calculates a dynamically changing surprise threshold. When the prediction error encountered by the ${M_t}$ exceeds the surprise threshold, the system enters into a search period to start using the best event model for predicting the following sensory input.

\subsubsection{Surprise threshold and search period}

The surprise threshold $\Phi_{M}$ is calculated by the rolling mean of the stored prediction errors $\overline{e}_{M}$ and of the variance $\sigma_{M}$ with a window (w) (see Figure ~\ref{fig:computational_model_process}B). The confidence parameter $\theta$ affects the tolerance of event models to the incoming errors and, thereby, controls the coarseness/length of the events to be segmented (see Figure ~\ref{fig:computational_model_process}C). $\Phi_{M}$ is calculated by

\[\Phi_{M} = \overline{e}_{M} + \theta * \sigma_{M}\]

If the prediction error of an event model exceeds the threshold $\Phi_{M}$, the system moves into a search period (see Figure ~\ref{fig:computational_model_process}D). Each search period starts with generating a new potential event model having random weights. Then, all event models in the system are trained for a rehearsal duration that indicates the number of training epochs. Following this training period, the best event model - the event model that receives the least mean squared error - is selected, and the effect of training on event models, except the best model, is ignored. Sometimes, the newly generated event model does not receive the minimum mean squared error, in this case, the new event model is removed from the long term memory.

During the search period, for each event, a different training set is generated to avoid catastrophic forgetting. For the prediction of the future, the next n timesteps in the sequence should be learned. Meanwhile, event models should not forget their history. Thereby, we sample a distinct training dataset for each event model from the combination of the next n timeteps and their histories.

\subsubsection{Memory range and replay}

At each search period an extensive search period takes place, which increases the training time as a function of the number of event models in the system. For this reason, at the end of each training epoch, we determine the unutilized event models to remove them when they are not utilized for n epochs (i.e., memory range). 

In order to foster memory consolidation, avoid catastrophic forgetting, and reduce the time spent for the training, we introduce a replay period in which each event model is trained by their histories (see Figure ~\ref{fig:computational_model_process}E). The replay period is a biologically plausible phenomenon observed in the hippocampal regions of the brain, which is thought to be related to the memory consolidation \citep{Olafsdottir2018Role}. It was also used for stabilizing the training regimes of the reinforcement learning agents \citep{Andrychowicz2017Hindsight}. 

Up to this point, we portrayed the formalization of the computational model. In the following section, we detail the preparation of the PLDs, which we used as the main stimuli for the psychological and computational experiments.

\subsection{Dataset and videos}

\begin{figure}[!ht]
\centering
  \includegraphics[scale=0.5]{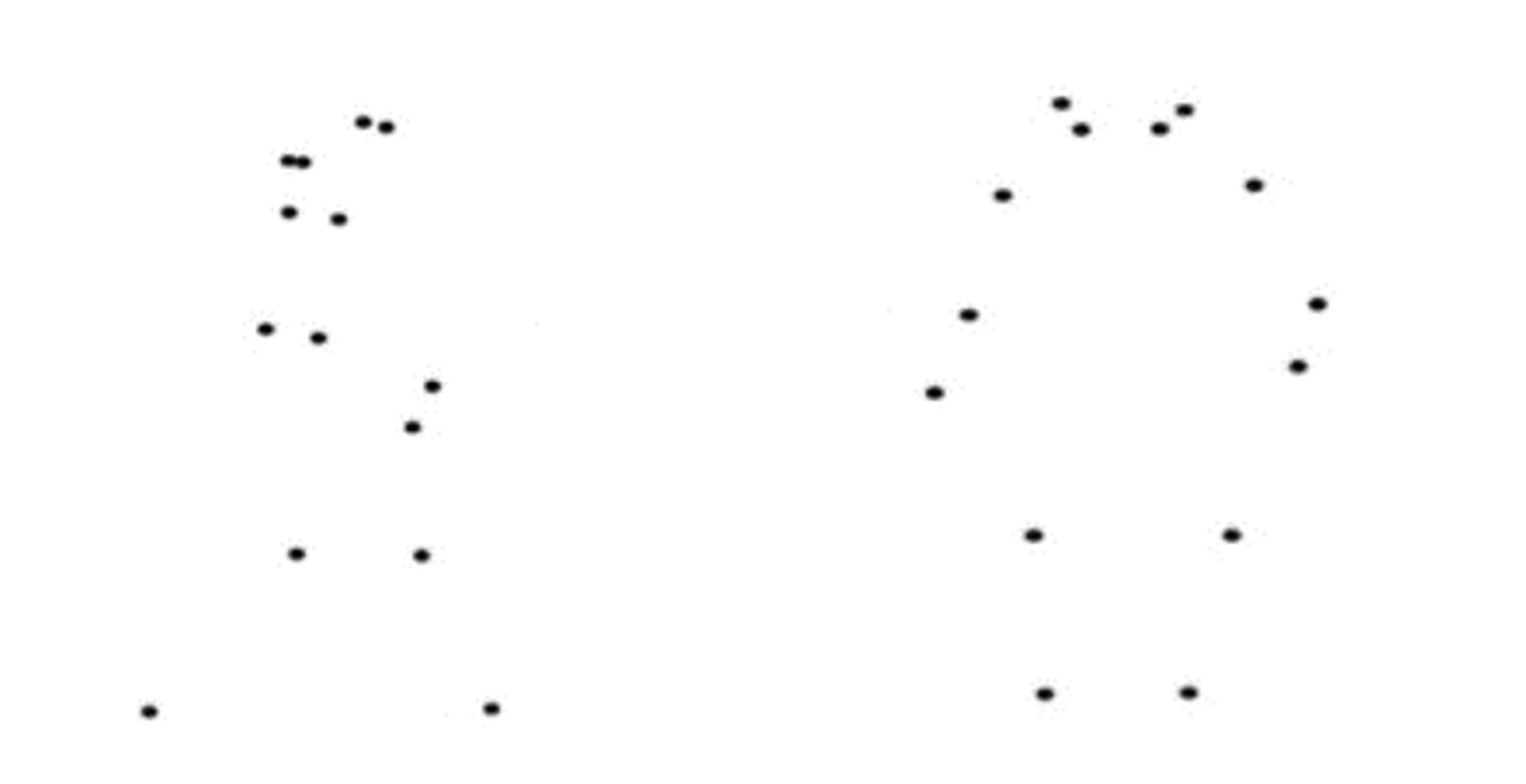}
    \caption{Two examples of PLDs are given. Both displays show a type of walking}
\label{fig:point_light_displays}
\end{figure}

For the preparation of the dataset from PLDs, we selected twelve natural human behaviors (such as walking, jumping, picking an object, sitting on a chair, and searching for an object) from the KIT Motion-Language Dataset \citep{Plappert2016KIT} (see Figure ~\ref{fig:point_light_displays}). The selected complex human behaviors were added back-to-back by an interpolation method (see Supplementary Material). All behaviors were represented in a PLDs format using the X and Y dimensions of 14 markers.

From the behaviors selected, we prepared two datasets/videos -normal and noisy- involving the same behaviors but in varying degrees of sensory change. By these datasets/videos, we assessed our model by examining its segmentation performance in comparison to the behavioral performance of humans, 

\subsubsection{Noisy dataset}

We produced the noisy dataset by exploiting the inherent relationship between the change and event segmentation \citep{Kurby2008Segmentation, Hard2011Shape, Zacks2007EventPerception, Zacks2020EventPerception, Cutting2012Perceiving, Hard2006Making, huff2014changes, Newtson1977Obj, Zacks2010Brain}. It is known that the event boundaries are the points where the change is notable the most and that the change is maximal particularly at the coarse-grained event boundaries \citep{Hard2011Shape}. That is, a reduction in change must hinder the potential timesteps that separate events from one another.

In order to prepare the noisy dataset, we applied Gaussian noise to the temporal signal of the normal dataset (window: 40, standard deviation: 10), and thereby, removed its fine movement dynamics and made it look much more fluent than the normal video. From normal and noisy datasets, we created two 16 Hz and 267-second videos for the psychological experiments. The videos can be found here \footnote{https://youtu.be/L2oCNEvzdrU}. For the computational experiments, we applied a min-max normalization operation to these datasets and reduced them to 4 Hz (1076 timesteps in total).

\section{Event segmentation experiments}

Humans can segment the ongoing activity into events in varying lengths. In the unitization paradigm, participants are asked to detect the shortest (fine-grained) and the largest natural/meaningful events (coarse-grained) \citep{Newtson1973Attribution, Zacks2007EventSegmentation, Zacks2007EventPerception, Hard2011Shape}. We conducted a unitization-based online psychological experiment that has two experimental conditions: event granularity (fine-grained and coarse-grained segmentation) and sensory reliability (normal and noisy videos). The number and positions of the event boundaries were determined to be the dependent variables.

In addition, we tested whether our computational model (1) captured the human event segmentation decisions and (2) showed a bias against the noisy video (i.e., the reduction of the rate of change) similar to that was expected in the performance of the human participants.

\subsection{Participants and procedure}
\label{subsection:event_seg_part_proc}

Nineteen participants (9 female, mean age 25) were recruited for a within-subject design. They were primarily undergraduate or graduate university students and were given the chance of a gift voucher lottery as an incentive. The study was approved by the Ethics Committee for the Master and PhD Theses in Social Sciences and Humanities (SOBETİK) at Boğaziçi University.

We reached the participants via the internet and, upon their agreement, sent them a link of the online experiment (i.e., Pavlovia) developed by Psychopy3 \citep{Peirce2019Psychopy}. Each participant segmented first the normal and then the noisy video. The experiment started with a segmentation granularity instruction which asked participants to make either a fine- or a coarse-grained segmentation by pressing the space button and the level of granularity was counterbalanced. For each type of segmentation granularity, participants observed the same video twice to segment the video either in the shortest or the longest possible way.

We coded each observation by a label for maintaining simplicity, namely Fine 1, Fine 2, Coarse 1, and Coarse 2. For example, Fine 1 denotes the first observation of the fine-grained segmentation level. Throughout our analyses we considered second observations as they were considered to be more reliable due to practice effect.

We trained twelve computational models for each possible combination of our independent variables. Please see the supplementary materials for the hyperparameter selection.   

\subsection{Results}

One challenge posed by the online psychological experiments is maintaining the data reliability \citep{Gosling2015Internet}. For this reason, we applied a number of control measures to ensure the quality of the data before applying the statistical tests. Firstly, we checked whether participants understand the instruction of sensory granularity by comparing their number of responses in coarse to those in fine segmentation condition. We excluded two participants as they had more number of responses in the Coarse 2 than the Fine 2 in the normal video segmentation. Secondly, we calculated the intra-participant correlations of the fine- and coarse-grained segmentation responses. This procedure resulted in the exclusion of three more participants (see Supplementary Material). At the end, 14 participants remained, whose data were considered for the statistical analysis.

\subsubsection{The reduction in sensory change reduces the number of event boundaries}

For analyzing the effect of the rate of change in fine- and coarse-grained event segmentation decisions, we used Friedman's ANOVA. Analysis revealed that the number of responses differed significantly across groups (X\textsuperscript{2}(7) = 80.93, \textit{p} = .000). Wilcoxon signed-rank was applied along with Holm–Bonferroni method to adjust the \textit{p} values to avoid inflating false positives. As a result, it was found that participants perceived a smaller number of event boundaries in the noisy (\textit{M} = 32.71, \textit{Md} = 17.5, \textit{SEM} = 4.44) than in the normal video (\textit{M} = 41.91, \textit{Md} = 22.5, \textit{SEM} = 5.95), \textit{W} = 879.0, \textit{p} = 0.021, \textit{r} = 0.191. The trend was similar for both fine-grained and coarse-grained segmentations. For the fine-grained segmentation, the normal video (\textit{M} = 68.14, \textit{Md} = 43, \textit{SEM} = 9.56) received more response than the noisy video (\textit{M} = 52.50, \textit{Md} = 43.5, \textit{SEM} = 6.98), \textit{W} = 288.5, \textit{p} = 0.016, \textit{r} = 0.31. Similar to the fine-grained, the coarse-grained segmentation level showed the same trend, normal (\textit{M} = 15.67, \textit{Md} = 16, \textit{SEM} = 1.49) and noisy video (\textit{M} = 12.92, \textit{Md} = 10.5, \textit{SEM} = 1.66), \textit{W} = 203.5, \textit{p} = 0.018, \textit{r} = 0.33, with a slightly higher effect size. These results verified that the reduction in sensory change in fact reduced the number of event boundaries, as expected. 
 
\subsubsection{Fine-grained event boundaries are more correlated with sensory change than coarse-grained event boundaries}
\label{subsubsection:Fine-grained-event-boundaries}

\begin{figure}[hbt!]
\centering
  \includegraphics[width=\columnwidth]{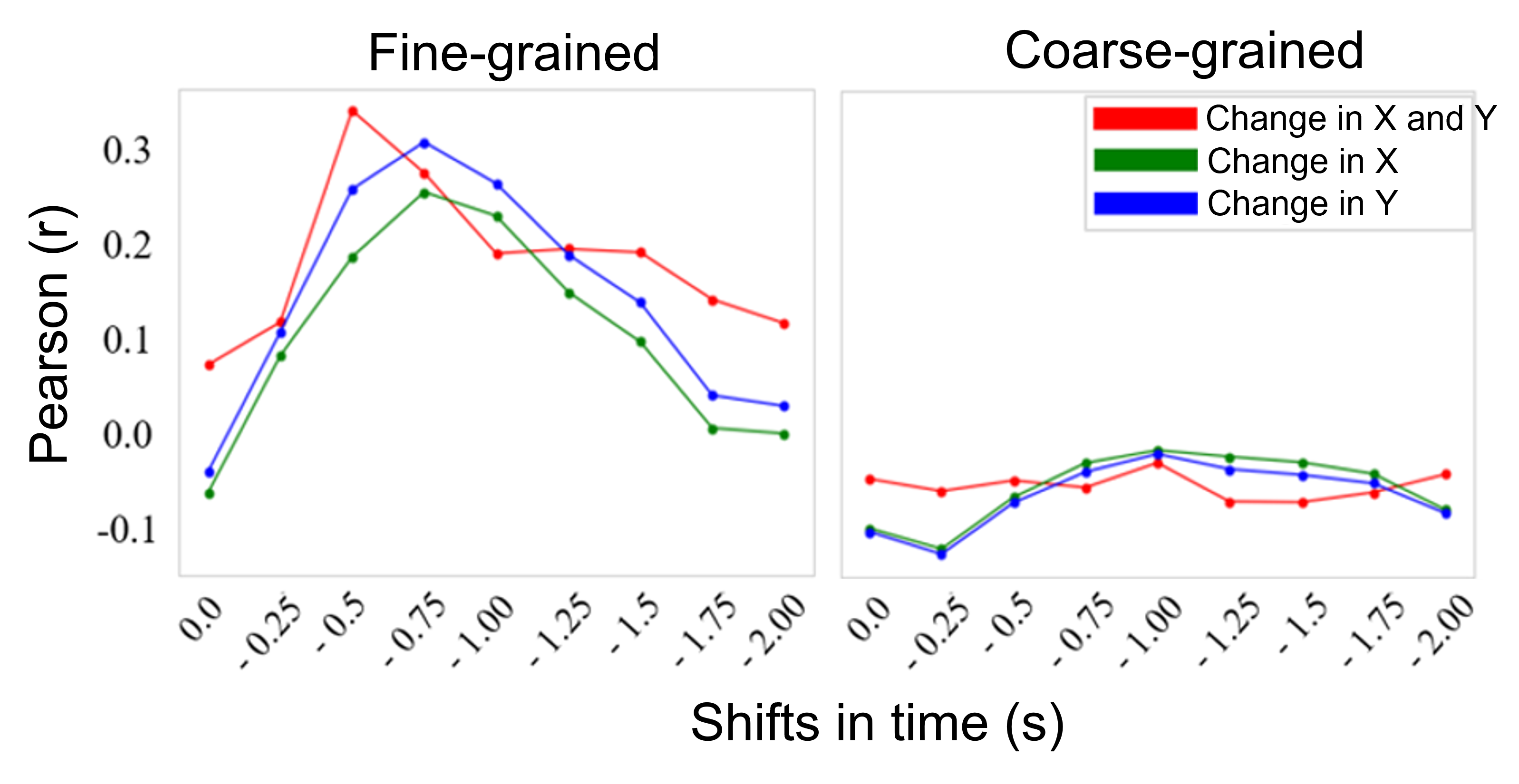}
    \caption{Event boundary responses and the degrees of change are given. Since participants detect event boundaries by pressing a button, there might be latencies in the responses. For this reason, response probabilities were calculated by event histograms and gradually shifted backward in time. For all the time shifts, the correlation between the response of probabilities of the participants and the absolute sensory change was computed. Results showed that fine-grained segmentation are correlated with the absolute sensory change, whereas coarse-grained segmentation decisions seem to be uncorrelated.}
\label{fig:change_and_responses}
\end{figure}

Aiming to reveal the relationship between sensory changes in point locations in two dimensions and event segmentation decisions of participants as a function of time, we generated histograms by grouping the responses of the participants into the 0.25-second bins and quantified changes as the absolute sum differences between the X and Y dimensions of points. Considering a possible lag between event boundary responses and changes in events, which may result from the decision-making process or the motor response latencies, we computed the correlations by shifting the response distributions backward in time. The relationship between the sensory change and the event segmentation are given in Figure ~\ref{fig:change_and_responses}. Figure ~\ref{fig:change_and_responses} shows that whereas the absolute sensory change is correlated with the rate of perceived fine-grained event boundaries; it is uncorrelated with the coarse-grained event boundaries. 

\subsubsection{Event boundaries of the model correlate with those of participants}
We received segmentation decisions of our computational model (Figure ~\ref{fig:computational_model_process}G). A video displaying fine- and coarse-grained segmentation instances can be found here \footnote[1]{https://youtu.be/SDQpmdehG8o}. After receiving the results, we compared them with the data of participants by point-biserial correlation (Figure ~\ref{fig:computational_model_process}I). Point-biserial correlation is a suitable technique for assessing the correlation between the event boundary responses (discrete) and the event boundary histograms (continuous) \citep{Franklin2020Structured}. For this reason, we generated a 1-second event boundary histogram for each condition and assessed event boundary responses of each participant and each model by this histogram. For testing whether the results of our model perform better than the chance level, we devised two control models: random and change models. While the former selects event boundaries uniformly by giving each timestep the same weight, the latter selects timesteps by different weights corresponding to their respective absolute sensory changes.

\begin{figure}[htp!]
\centering
  \includegraphics[width=\columnwidth]{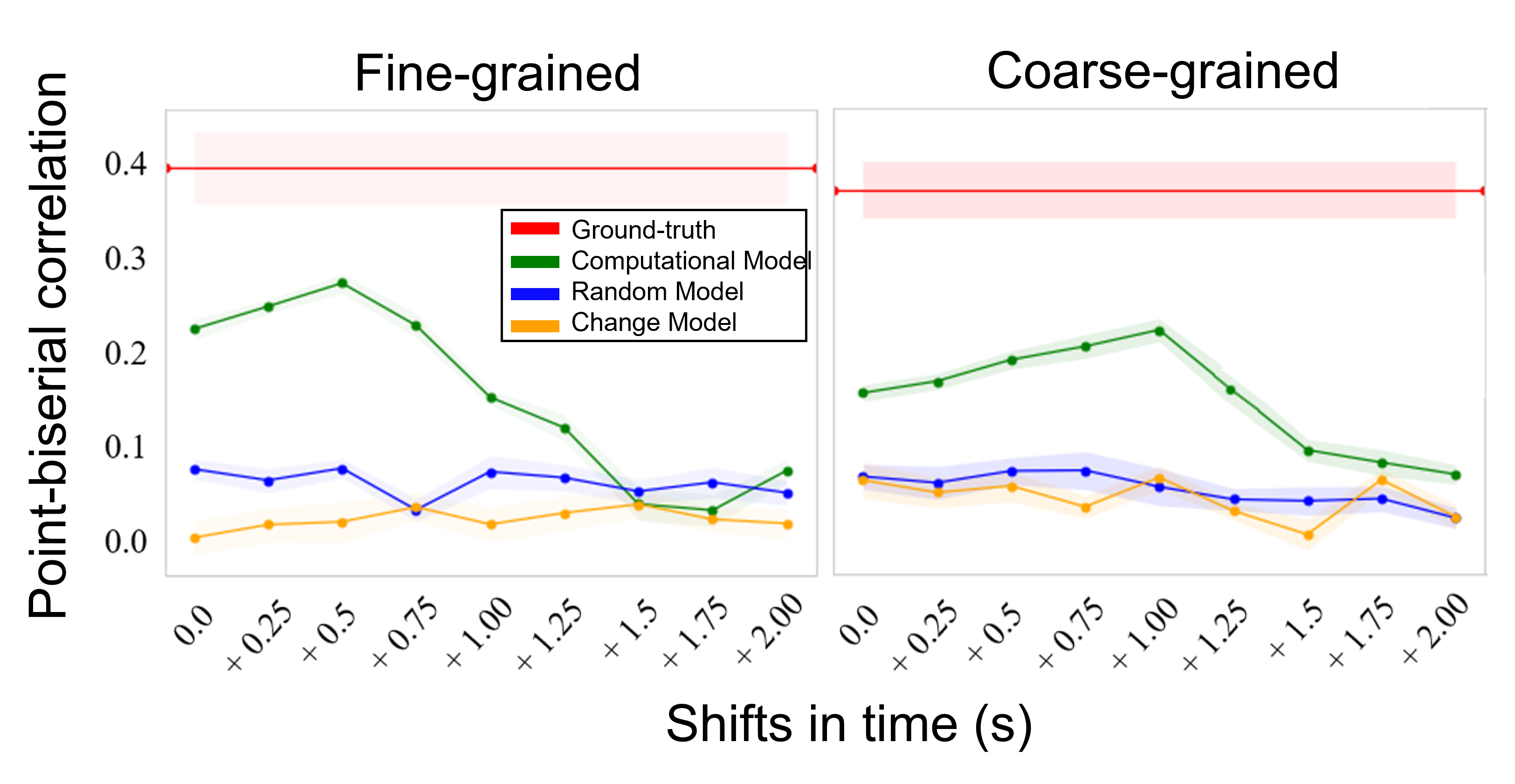}
    \caption{Time-dependent correlations of the computational models with the ground-truth data are given. In the unitization paradigm, participants might show a response latency. Considering this fact, we shifted the responses of the models forward in time before computing the point-biserial correlations. Our model reached better correlation scores with the human performance than the control models.}
\label{fig:normal_model_performance}
\end{figure}

Directly comparing the performances of humans and computational models may fail to reveal the actual match because models can respond to an event change right away while humans show response latencies. Due to this fact, for a thorough comparison, we assessed the performance of models by shifting their responses in time. In details, we calculated the point-biserial correlation values with the temporal shifts up to 2 seconds (8 frames) (see Figure ~\ref{fig:normal_model_performance}).

For the fine-grained event segmentation, our model reached its highest performance at a 0.5-second shift with an \textit{r} value of 0.254 when the mean \textit{r} of the ground-truth data was 0.394. The coarse-grained event segmentation performance of our model was maximized after a 1-second shift with an \textit{r} value of 0.196, when the mean of participants was \textit{r}\textsubscript{mean} = 0.366. Although this is a significant difference, the particular model receiving maximum correlation scores (\textit{r}\textsubscript{max} = 0.256) showed a comparable performance to those of participants (\textit{z} = -1.40, \textit{p} = 0.158, two-tailed). In general, these results showed that segments produced by our model correlate with those produced by human observers.

\subsubsection{Event boundaries produced by the computational model correlate with sensory change}
\begin{figure}[!ht]
\centering
  \includegraphics[width=\columnwidth]{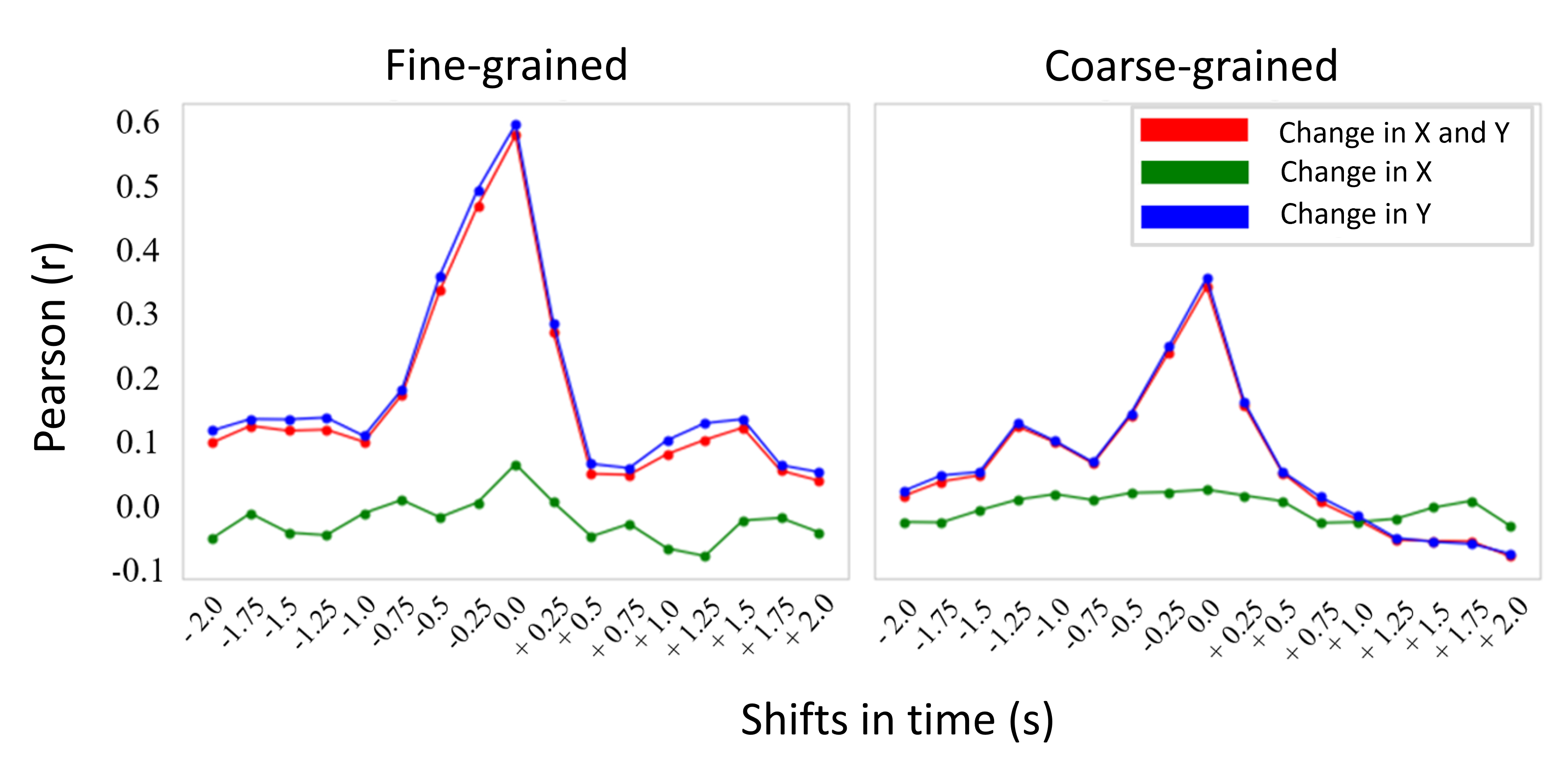}
    \caption{The relationship between the responses of the proposed model and the change is given. Time-dependent correlations between the absolute sensory change and the responses of the  computational models for the fine-grained and coarse-grained segmentation. Coarse-grained segmentation decisions of the model were less correlated with the absolute change than its fine-grained segmentation decisions. In comparison to the participants, the computational model reacts to the change immediately.}
\label{fig:change_and_model_responses}
\end{figure}

We also investigated the relationship between the sensory change and the responses of the computational models as showed in Figure~\ref{fig:change_and_model_responses}. Results showed that the fine-grained segmentation decisions of the computational models were more accountable by the absolute sensory change than its coarse-grained segmentation decisions. In this sense, this pattern is roughly analogous to the asymmetry emerging in the responses of the participants (see Figure ~\ref{fig:change_and_responses}). 

\subsubsection{Sensory noise does not bias the computational model similar to participants}
\label{subsubsection:Sensory-noise-does-not-bias}
\begin{figure}[!ht]
\centering
  \includegraphics[scale=0.25]{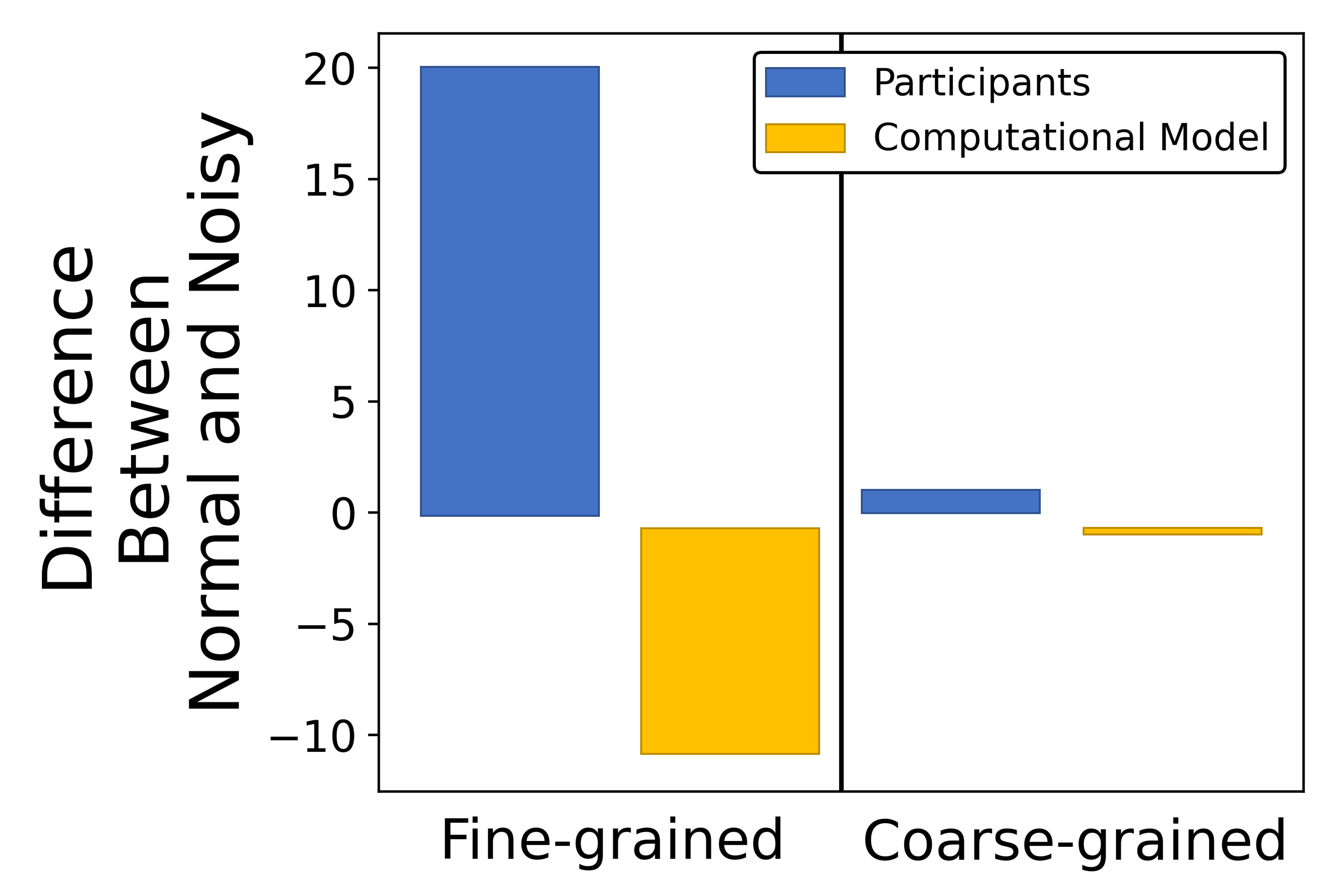}
    \caption{The figure shows the difference between the mean number of event boundaries detected in the normal and noisy videos by the participants and computational models. The computational model did not show a similar bias in the fine-grained segmentation, but the results were comparable in the coarse-grained segmentation.}
\label{fig:model_bias}
\end{figure}

We have so far showed that, for the normal video, event segments produced by our model correlate with those produced by human observers. We further tested our model to check whether it would be biased by the quality of sensory input, in other words, by the reduction in the sensory change. In participants' data, the number of segments produced for the noisy video was less than those produced for the normal video. Figure ~\ref{fig:model_bias} shows the number of event boundaries produced by the participants and our computational model. The model did not show a similar bias, although the event boundary decisions results were comparable.

\hamit{In the first part of this study, we investigated how the reduction of sensory change affects event boundaries detected by participants. Results showed that fine-grained event boundaries of participants correlated with sensory change, whereas coarse-grained ones did not (Section~\ref{subsubsection:Fine-grained-event-boundaries}). Additionally, when the sensory information was degraded, the number of fine-grained event boundaries detected by participants were reduced. Then, we assessed our computational model and found out that decisions of our model were correlated with participants and its fine-grained segmentation decisions showed higher correlation with sensory change than its coarse-grained segmentation decisions. Lastly, we checked whether our model was biased by degraded sensory change similar to participants by comparing their number of event boundaries in normal and noisy videos. Our model did detect higher number of fine-grained event boundaries for the normal video, which was contrary to our expectations (Section~\ref{subsubsection:Sensory-noise-does-not-bias}).}

\hamit{Overall, in this section, we showed that our model captures the essence of event segmentation by exhibiting similar event segmentation decisions except showing a bias to the noise in sensory information. In the second part, we assessed the internal event representation space and the corresponding similarity judgments of the model with those of humans.}

\section{Event representation experiments}

The aim of this section is to analyze and compare the event representations formed by the human participants and our computational model. However, representations formed by humans are not directly accessible. Therefore, using the observed PLD streams and the corresponding ground truth event labels, two metrics are calculated in human experiments:
\begin{itemize}
    \item {\bf Pairwise Event Distance in Humans (PED-H):} Assuming that events with similar representations are reported to be similar by humans, we utilize similarity judgements of the participants to collect pairwise event distances in humans. 
    \item {\bf Estimated Event Representation in Humans (EER-H):} Given pairwise distances between each event, we estimate the position of each event in a reconstructed representation space using non-linear dimensionality transformation techniques. This transformation places each event instance in a 2-D event representation space such that the pairwise distances between events are preserved.
\end{itemize}

Taking into account the metrics obtained in human experiments above, the following two metrics, which serve as the counterparts in our computational model, are used:
\begin{itemize}
    \item {\bf Pairwise Event Distance in Computational Model (PED-CM):} Recall that each event in our model is represented with a feed-forward neural network (FFNN). The pairwise distances between events are calculated by taking the distances between the outputs of the corresponding event models.  
    \item {\bf Estimated Event Representation in Computational Model (EER-CM):} 
    We train an event model classifier, which is another neural network that maps the PLD sensory data stream to the corresponding event model id ($M_j$). Here, the neural activations, which are generated by the PLD data streams, encode the representations of the corresponding events. These activations are mapped into 2-D event representation space using dimensionality reduction techniques. The pairwise distances between events are calculated by taking the distances between their locations in 2-D event representation.
\end{itemize}

Finally, we propose two additional baselines to assess metrics defined for the computational model: PED-CM and EER-CM. These metrics are not computed from the behavior of the computational model or the similarity judgments of humans. They are computed based on the ground-truth data, in other words, the visual appearance of PLDs streams. We expect this assessment to show whether the defined metrics and, therefore, our computational model capture distances between and representations of events similar to or better than ground-truth data.

\begin{itemize}
    \item {\bf Pairwise Event Distances based on Visual Appearances (PED-VA):} The pairwise distances between events are estimated based on their visual appearances. 
    \item {\bf Pairwise Event Distance based on Changes in Visual Appearances  (PED-CVA):} PED-VA is based locations of dots, which are necessarily dependent on previous dot configurations in an event. Thus, for estimating distances between events without the bias of previous dot locations, we estimated distances by changes in visual appearances of events.
\end{itemize}

In the rest of this section, we first describe the details of the human similarity judgment experiments, next provide the computation details of the metrics introduced above, and finally give the results.

\subsection{\it Procedure}

\subsubsection{Event Similarity Judgment Experiment with Human Participants}

We devised an online psychological experiment based on the pairwise similarity judgment paradigm \citep{Shepard1980Multidimensional, Shepard1987Toward, Shepard1979Additive} (see Figure~\ref{fig:computational_model_process}J). In this paradigm, participants are asked to report the similarity between two items, and the obtained similarity matrices are transformed to a representation space by multidimensional scaling \citep{Shepard1980Multidimensional}.

In our experiment, participants were presented with the videos side-by-side and had a chance to play videos using the K and L buttons (K for left, L for right) of the keyboard. They reported the degree of similarity of event videos in both fine- and coarse-grained levels using a continuous ranking slide. Psychological experiments were designed by Psychopy3 \citep{Peirce2019Psychopy}, and online experiments were conducted via Pavlovia.

Since the number of segments increases the number of possible comparisons, we reduced the number of selected events by sampling the ones that were also segmented by our computational model (see Section 4.1.2 and Supplementary Material for details). This sampling schema yielded 21 videos for the fine- and 12 videos for the coarse-grained segmentation. In order to further decrease the number of comparisons, we randomly selected 7 fine-grained and 4 coarse-grained event videos for each participant. This reduced the number of comparisons to 49 for the fine- and 16 coarse-grained videos.

Forty-two participants (32 female, mean age 21) were recruited for the study from the Research Participation System of Boğaziçi University. Other details were same as those in the event segmentation experiment (Section~\ref{subsection:event_seg_part_proc}).

This event similarity judgment experiment provided us with PED-H, the similarity scores between the events in both fine- and coarse-grained levels. We calculated EER-H metrics from PED-H by multi-dimensional scaling in the scikit-learn package \citep{scikit-learn}.

\subsubsection{Event Similarity Judgment Extraction from the Computational Model}

We defined two metrics for comparing representations of computational model with those of participants, namely PED-CM and EER-CM.

The PED-CM is calculated by taking the distance between the outputs of the corresponding event models. Recall that each event model in our computational model is represented with a learned feed-forward neural network which predicts the PLD configuration in the next time step given the displayed PLD configurations. If the learned models are similar, then we expect similar outputs, i.e. prediction trajectories, from those models. Therefore, given arbitrary PLD streams, the distance between the prediction trajectories of two event models provides a measure for their pairwise similarity. In order to calculate the distance between two prediction trajectories, Dynamic Time Warping (DTW) method \citep{Giorgino2009Computing} is used (Supplementary Material for details).

EER-CM, on the other hand, is calculated by training a neural network classifier that takes PLD data stream as input and predicts the corresponding event model id as output. The neural activations of this classifier encodes the predicted event by our computational model (see Figure ~\ref{fig:computational_model_process}H). These activations are mapped to 2-D representation space using dimensionality reduction methods such as PCA (Principal Component Analysis) \citep{wold1987principal} or t-SNE (T-Distributed Stochastic Neighbor) \citep{van2008visualizing}. Finally, the mean location of the points that belong to the same event in the 2-D representation space is found and assigned as the representative position for the corresponding event. The distance between events, on the other hand, are found by calculating the Euclidean distance between these representative positions. Due to the two dimensionality reduction methods, we received two different metrics for EER-CM by using PCA (EER-CM/PCA) and using t-SNE (EER-CM/t-SNE).

\subsubsection{Similarity Judgments based on Visual Appearances}

We invented two additional baselines for the evaluation of similarity judgments and estimated event representations of the computational model, namely PED-VA and PED-CVA. These techniques are used to estimate pairwise visual similarities between events. PED-VA estimates visual similarities based on actual trajectories of events; on the other hand, PED-CVA uses changes in trajectories of events. In order to calculate the distance between actual trajectories and their changes, Dynamic Time Warping (DTW) method \citep{Giorgino2009Computing} is used.

In this part of the study, we explained how we extracted several metrics for the participants, the computational model, and baselines. Our aim is to find out (1) whether similarity is a meaningful metric for events and (2) whether our model captures event similarity judgments, in other words, estimated event representations of participants. In particular, the performance of our model in this task would be a sign for validating our model.

\subsection{\it Results}

In this section, we compared subjective similarity judgments of the participants with the metrics we calculated for the computational model and the baseline. In particular, first we assessed the correlation between the similarity judgment of participants, namely PED-H, with those of the computational model (PED-CM and EER-CM) and the baselines (PED-PLDs and PED-CPLDs) in Section \ref{similarity-judgments-correlation}. Next, we visually compared event representations of participants (EER-H) with representations estimated from the computational model (EER-CM) in Section \ref{som-visual-similarities}.

\subsubsection{Event similarity judgments of participants show strong correlation to their group}

We averaged out the subjective similarity judgments of the participants to compute the group similarity judgments (PED-H) and compared the performances of the individual participants with the group decisions by using Pearson correlation coefficients. This investigation (1) revealed correlation distributions of the participants and (2) provided a basis for assessing the representation discovery techniques. Results showed that that the mean Pearson correlation coefficients of the participants to their group were 0.90 and 0.92 for the fine- and coarse-grained event comparisons, respectively. 

\subsubsection{Event similarity judgments of the computational model and participants correlate each other} \label{similarity-judgments-correlation}
After receiving the group similarity judgments for event pairs (PED-H) from the event similarity judgment experiment, we compared them to similarity judgments we computed for the computational model and the baselines by Pearson correlation. Recall our metrics for the computational model, namely PED-CM and EER-CM, for the representation discovery. We received two results from the EER-CM because of using two-dimensionality reduction techniques (EER-CM/PCA and EER-CM/t-SNE) and one result from the PED-CM. For the baselines, we received two results; one for the PED-PLDs and one for the PED-CPLDs.

\begin{figure}[!ht]
\centering
  \includegraphics[scale = 0.30]{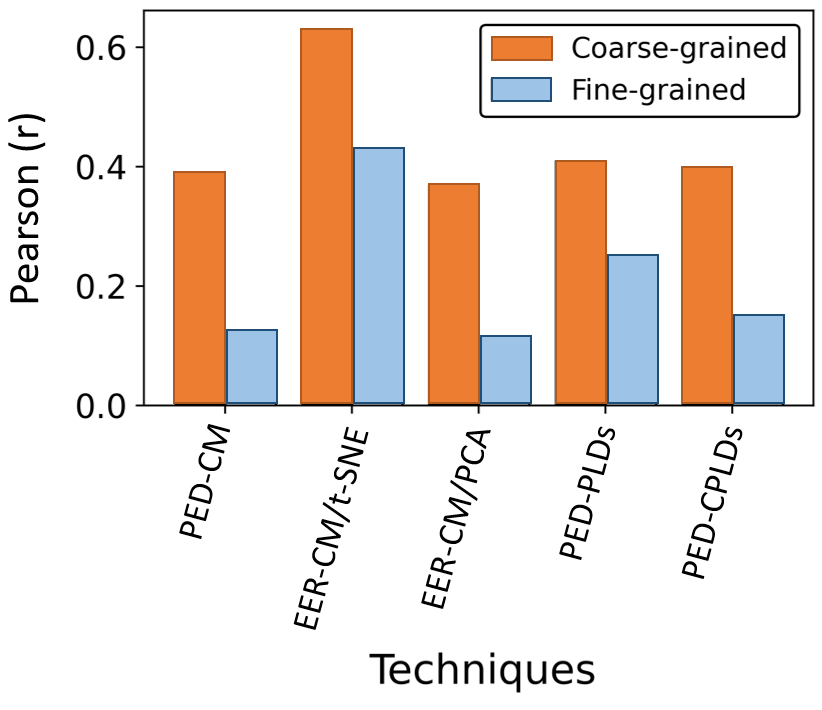}
    \caption{Correlations of the representation discovery techniques and baselines with the human data are given. The best technique was EER-CM/t-SNE receiving an outstanding performance compared to the baseline models, namely PED-PLDs and PED-CPLDs. Even though the correlation scores were not approximated perfectly, they were comparable to the human performance, namely 0.90 and 0.92 for the fine- and coarse-grained event comparisons, respectively.}
\label{fig:event_representation_discovery_techniques_correlation}
\end{figure}

Figure ~\ref{fig:event_representation_discovery_techniques_correlation} displays the correlations of these metrics with group decisions. It can be seen from the figure that the EER-CM/t-SNE was better than other techniques and baselines (\textit{r} = 0.435 for fine-grained and \textit{r} = 0.614 for coarse-grained). Statistical tests revealed that correlation scores of EER-CM/t-SNE (\textit{r} = 0.618) with PED-H was higher than EER-CM/PCA (\textit{r} = 0.393), (\textit{z} = 2.579, \textit{p} = 0.004) for coarse-grained units. Moreover, the same also applied for fine-grained units where EER-CM/t-SNE (\textit{r} = 0.449) was significantly higher than EER-CM/PCA (\textit{r} = 0.339), (\textit{z} = 1.92, \textit{p} = 0.027). 

These results suggested that the metric estimating event representations in the computational model by a nonlinear transformation technique (EER-CM/t-SNE) showed better results than the other techniques, even superior to baselines estimating pairwise event distances based on visual appearances (PED-VA and PED-CVA). That is, metrics based on visual appearances cannot capture human similarity judgments (PED-H) well. We showed that, for capturing human similarity judgments, a higher-level mechanism (i.e., event model classifier) mapping sensory information (i.e., PLDs in the current case) to event units seems necessary. In fact, in the literature, a higher-level mechanism capturing the relationship between event pairs is proposed \citep{aslin2017statistical, levine2019finding}.

Although a higher-level mechanism for capturing similarity judgments might be necessary, it is apparent that it seems it is not sufficient. Recall that EER-CM/PCa computes similarities between events based on the activations of a higher-level mechanism. Despite this, EER-CM/PCA does not correlate well with human similarity judgments (PED-H), and it seems inferior to baselines based on visual appearances (PED-VA and PED-CVA). The dimensionality technique is the only difference between EER-CM/t-SNE, showing superior performance by using a nonlinear dimensionality reduction technique \citep{van2008visualizing}, and EER-CM/PCA, showing inferior performance by using a linear dimensionality reduction technique \citep{wold1987principal}. Superiority of EER-CM/t-SNE over EER-CM/PCA suggests that the former can represent the relationships of events in higher dimensions on a 2-D event representation space by nonlinear transformations. Thus, we can conclude that the higher-dimensional space formed by the event model classifier correlates with those of humans.

So far, we have demonstrated that (1) people can judge similarities between events, and (2) our model shows remarkable correlation scores with the human similarity judgments. 

For a visual exploration of event representations beyond correlation scores, we compared EER-CM/t-SNE with EER-H on a self-organizing map (SOM) \citep{Kohonen1990SelfOrganizing}. SOM is an unsupervised machine learning technique that produces the low-dimensional representation of data. In the literature, SOM is used is used for action recognition \citep{huang2010human, gharaee2021online}, classification \citep{wunstel2000behavior} and analysis \citep{bauer1997self}. Moreover, compared to t-SNE and PCA, SOM preserves topological relationships in the data \citep{dou2021v}, which makes it suitable for the visualization of events represented by EER-CM/t-SNE on a 2-D space.

\subsubsection{Self-organizing maps reveal visual similarities between represented events}
\label{som-visual-similarities}

\hamit{SOM is an unsupervised learning technique that extracts the structure of the data by preserving topological relationships \citep{dou2021v, Kohonen1990SelfOrganizing}. Upon training, SOM forms a topographic map according to the patterns in the input data and represents statistical patterns by the distances between elements \citep{Kohonen1990SelfOrganizing}. In this part, we used SOM models to visually explore event representations of the computational model and the participants on the same 2-D space.}

\hamit{To visualize fine- and coarse-grained events on a 2-D space \citep{dou2021v}, we trained two SOM models for both the computational model and the participants. SOM models trained for participants received their input from the PED-H that is gathered from the event similarity judgment experiment, whereas those of the computational model received their input from EER-CM/t-SNE as it reached the best correlation score. We thereby trained SOM models by representing each event by its perceived (i.e., the participants) or estimated (i.e., the computational model) distance to other events in the dataset.}

\hamit{As a result of the training, we obtained locations of events for the computational model and the participants on different 2-D spaces. To make the comparison more accessible, we mapped the event locations of the computational model to the 2-D space of participants. For mapping, we used the orthogonal Procrustes method \citep{Schonemann1966GeneralizedSolution} in Scipy \citep{2020SciPy-NMeth} since it preserves pairwise distances between events on a 2-D space.}

\begin{figure*}[hpbt!]
\centering
  \includegraphics[scale=0.8]{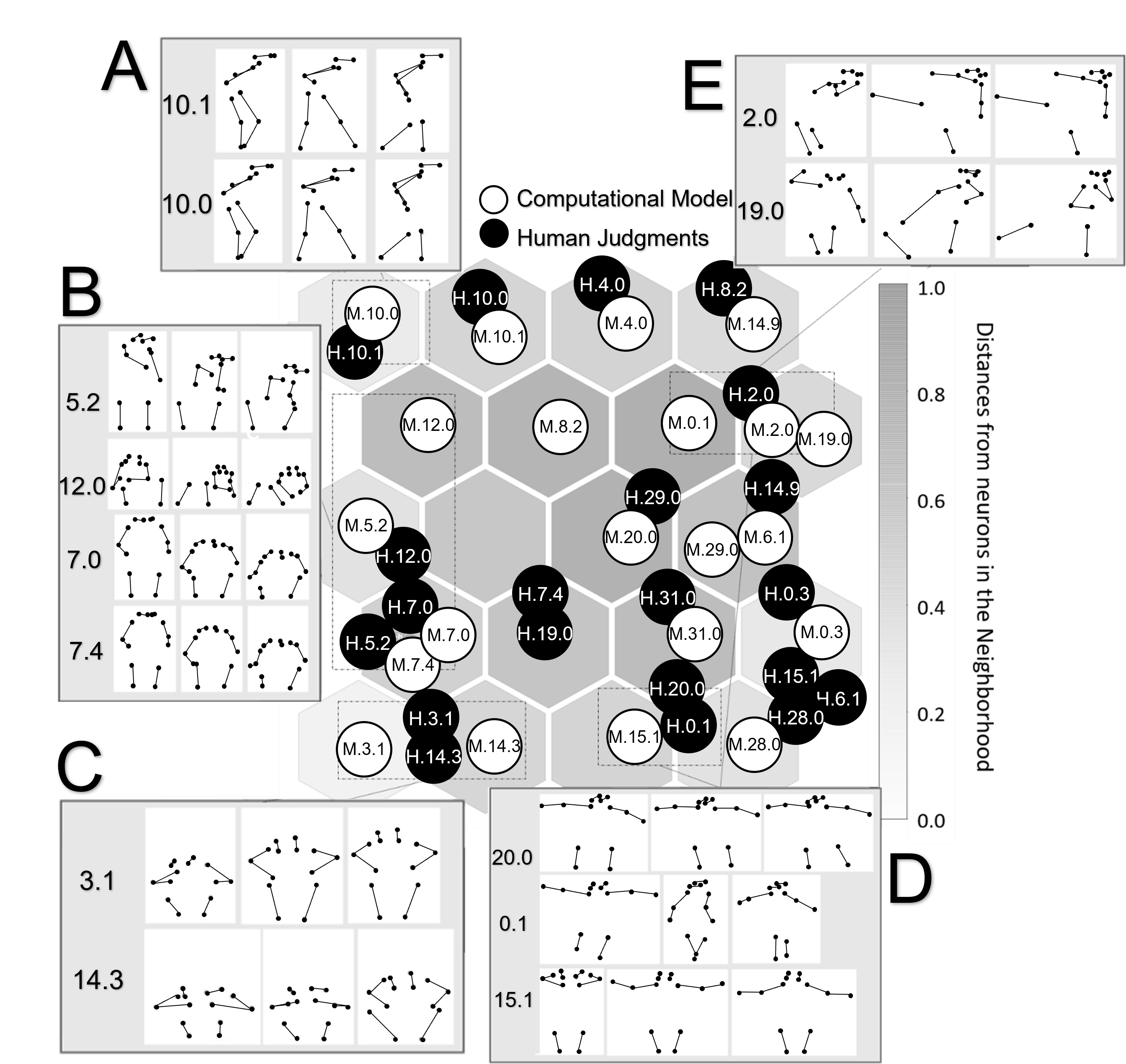}
    \caption{SOM results for the fine-grained event representations are given. Event segments, in general, are represented by the same or neighborhood neurons (see 10.0 and 10.1, 7-0 and 7.1). Moreover, SOM extracted the topology, as expected. For example, (A) represents walking, (B) walking and leaning down, (C) a complete bending (i.e., crouching down), (D), and (E) hand movements. Despite this, there are several deviations. For instance (D and E), the computational model considered 0-1 more similar to 2.0 and 19.0, whereas humans thought it was more like 20.0. Lines connecting the points are pictured for visualization purposes.} 
    
\label{fig:som_fine_grained_segmentation}
\end{figure*}

The resulting 2-D space representing the fine-grained events was given in Figure~\ref{fig:som_fine_grained_segmentation}. The hexagons represent neurons of the SOM model on which events of the computational model and the participants are located, and colors of hexagons represent their distances from surrounding neurons (i.e., hexagons). Events are labeled by a letter and two numbers, such as M.0.7. The letter shows whether the event is represented by the model (M) or humans (H), the first number indicates the event model used for the event, and the second number shows the order of the segment that particular event model was used for. For a visual comparison, we illustrated PLDs streams corresponding to events and organized them in different boxes. For example, Figure~\ref{fig:som_fine_grained_segmentation}A includes two events labeled as 10.0 and 10.1, which are represented by neighboring neurons. The spatial closeness on the 2D space represents the degree of similarity. Throughout this section we will evaluate whether this similarity corresponds to the visual and the semantic similarities between event segments.

It can be seen that the representations of the computational model and participants were well-matched to one another (see event segments coded as 10.0 and 10.1 in Figure~\ref{fig:som_fine_grained_segmentation}-A, and 7.0 and 7.1 in Figure~\ref{fig:som_fine_grained_segmentation}-B), aligning with the resulting Pearson correlation coefficient (r = 0.435). Moreover, SOM displayed the topological relationships between the represented units. For example, Figure~\ref{fig:som_fine_grained_segmentation}-A represents walking, Figure~\ref{fig:som_fine_grained_segmentation}-B walking and leaning down, Figure~\ref{fig:som_fine_grained_segmentation}-C a complete bending (i.e., crouching down), Figure~\ref{fig:som_fine_grained_segmentation}-D and Figure~\ref{fig:som_fine_grained_segmentation}-E various kinds of hand movements. 

\hamit{Despite matches between representations of the computational model and the participants, there are several disagreements (see event segment coded as 0.1 (see Figure~\ref{fig:som_fine_grained_segmentation}-D and Figure~\ref{fig:som_fine_grained_segmentation}-E). The computational model considered 0.1 more similar to 2.0 and 19.0, whereas participants perceived it as more like 20.0. Interestingly, the neuron representing M.0.1 is distant from surrounding neurons (see the color of the hexagon), which implies that the computational model could not estimate the similarity of M.0.1 with other events properly and assigned it to an individual neuron.}

\begin{figure*}[hpbt!]
\centering
  \includegraphics[scale=0.75]{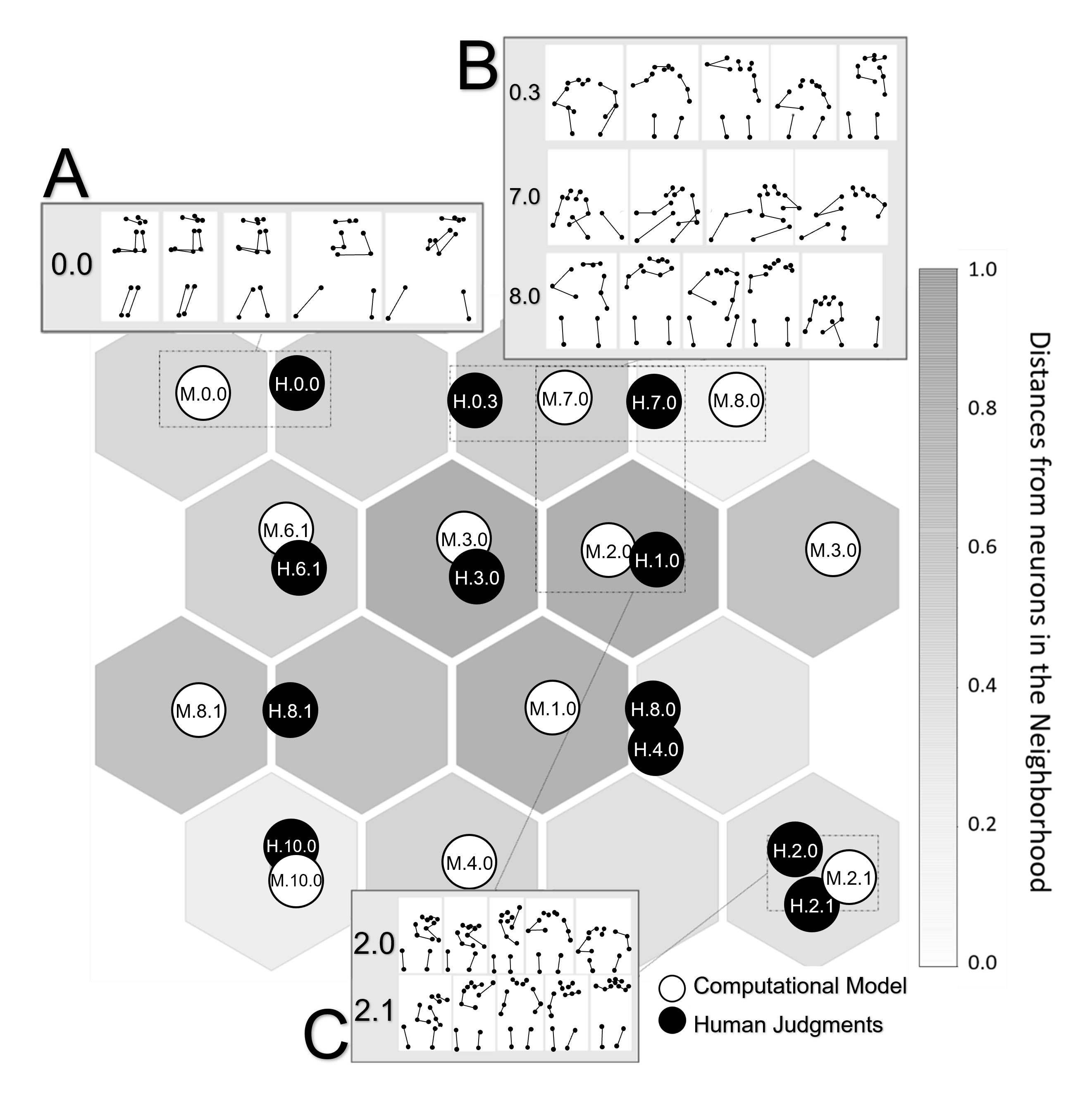}
    \caption{SOM results for the coarse-grained event representations are given. (A) shows an instance, (B) shows the relationships between three coarse-grained events, namely 0.3, 7.0, and 8.0. They share a similar movement, which can be defined as bending. On the other hand, (C) shows a divergence between humans and the computational model. Humans, but not the computational model, found 2.0 (raising one hand) and 2.1 (raising two hands together) similar. Lines connecting the points are pictured for visualization purposes.}
\label{fig:som_coarse_grained_segmentation}
\end{figure*}

\hamit{Representations produced for the coarse-grained event segments were given in Figure~\ref{fig:som_coarse_grained_segmentation}. Similar to the fine-grained event segments, representations of the computational model and participants overlapped one another (see M.0.0 and H.0.0 and Figure~\ref{fig:som_coarse_grained_segmentation}-A). The overlap of representations is consistent with the degree of correlation scores received in similarity judgments (\textit{r} = 0.614). Since the coarse-grained event units represent rich and diverse movement sequences, interpreting their topological relationships is rather hard.}

\hamit{Event units may share similar sequences, and the computational model and the participants seem to use those similarities in their judgments. For instance, Figure~\ref{fig:som_coarse_grained_segmentation}-B displays three event segments: 0.3, 7.0, and 8.0. These event segments differ from one another, but they involve a similar sequence, a bending movement. The participants and the computational model might have spotted the shared movement between sequences and considered them similar (see M.7.0 and H.7.0 and compare them with M.8.0). This implies a shared strategy between the model and the participants.}

\hamit{Figure~\ref{fig:som_coarse_grained_segmentation}-C displays a crucial difference between the model and participants (see 2.0 and 2.1). Participants found 2.0 and 2.1 similar (see H.2.0 and H.2.1) as they involve nearly the same movement sequence. On the other hand, the model could not perceive them as similar (see M.2.0 and M.2.1). We think this might have been because 2.1 and 2.0 differed in one movement sequence, namely, raising both hands together, which took place in 2.1 but not in 2.0. It is natural that participants found raising one or two hands similar, but the model could not detect this fact. Although our model has a high correlation with human similarity judgments  (\textit{r} = 0.614) and produces overlapping representations on 2-D space, it fails to capture the relationship between events in certain cases because of failing to conceptualize the relationship between body parts (i.e., left and right hand) or actions.}

\hamit{In this part, we visually compared representations of the computational model and the participants to reveal their similarities and differences. In general, our model produced representations that overlap with those of participants. However, we detected several cases of failure due to not being able to estimate the similarity of an event and conceptualize the semantic relationships between body parts or actions. In the next section, we will discuss our results in the light of event segmentation and representation literature and present possible improvements over the proposed model.}

\section{Discussion}

In this study, we developed a computational model of event segmentation, learning, and representation by considering the limitations in the literature. In the literature, several studies used stimuli which have discontinuities and unnatural abruptions \citep{Reynolds2007Computational, Metcalf2017Modelling}, some only worked in the context of interaction \citep{Gumbsch2016Learning, Gumbsch2017ComputationalModelDyn, Gumbsch2019Autonomous}, and the majority of studies did not validate their model by psychological experiments \citep{Gumbsch2016Learning, Gumbsch2017ComputationalModelDyn, Gumbsch2019Autonomous, Reynolds2007Computational, Metcalf2017Modelling}. To the best of our knowledge, the only model validated by the ground-truth data was tested only for one granularity level \citep{Franklin2020Structured}. We addressed these limitations by developing a simple semi-mechanistic and self-supervised computational model and demonstrated that our model can produce multimodal segments in varying hierarchies.

\subsection{Humans and computers differ in terms of their temporal dynamics}

One interesting output of this study is that humans and computers are different in terms of their temporal dynamics, and researchers investigating the cognitive and perceptual abilities should be cautious about these differences \citep{Funke2021FivePoints}. In this study, we observed a delay in the event segmentation responses of participants, which might have occurred due to the decisional or motor processing delays. For a safe and sound comparison between the proposed model and the ground-truth data, we shifted the responses of the model in time. Our analysis showed that the proposed model received comparable mean correlation scores with participants, namely 0.254 and 0.196 for the fine- and coarse-grained segmentations, respectively, where the corresponding values received by the best models were 0.32 and 0.256. Steady improvement of the correlation scores as the time shifted forward implied that the comparison between humans and computational models is not straightforward. Nonetheless, shifting responses in time was an approximation; thus, further computational models can benefit from other paradigms of event segmentation; for example, the dwell-time paradigm can address to this limitation \citep{Hard2011Shape}. In the dwell-time paradigm, participants are presented with a video frame by frame and are asked to use a button for passing into the next frame. By considering the time spent for each frame, an event boundary histogram can be extracted without or with considerably less delay \citep{Hard2011Shape}. In comparison to the study of \cite{Franklin2020Structured}, which is the only study which validates the model with human response, our model received a higher correlation score. However, while \cite{Franklin2020Structured} used naturalistic videos directly, here, we used PLDs extracted from natural videos. Despite this fact, our simple model extracting the event boundaries by tracking the prediction error signals can also be extended to the naturalistic videos.

\subsection{Changes predict fine- but not coarse-grained event segmentation responses}

We conducted a series of comprehensive correlational analyses to reveal the relationship between the change and the event segmentation performances of the participants and our computational model (see Figure~\ref{fig:change_and_responses}). In general, we showed that the fine-grained segmentation decisions of humans were partly driven by the absolute sensory change. In contrast, their coarse-grained segmentation decisions seemed to be independent from the change in the sensory information. Analyzing the relationship between the event segmentation decisions of participants and the sensory change gave us a chance to interpret the capabilities of our model. Figure~\ref{fig:change_and_responses} shows that the fine-grained segmentation decisions of our model are largely correlated with the absolute sensory change. On the other hand, Figure~\ref{fig:change_and_model_responses} shows that the effect of absolute sensory change on the decisions of our model is reduced for the coarse-grained segmentation. This implies the possibility of a weak top-down effect developed by our model. 

In the second condition of our event segmentation experiments, where the videos to be segmented are noisy, we benefited from the well-known relationship between the change and the event segmentation. We expected that the reduction in sensory change and removing the fine-grained dynamics would affect the event segmentation decisions of our model in line with those of the participants. By verifying the relationship between the event segmentation and the sensory change, participants produced coarser segments, as expected (see Figure~\ref{fig:model_bias}). When we tested our model trained on the normal video for the noisy video, the number of boundaries did not show the bias we expected. We think that the noise applied to the dataset might have increased the prediction error signals observed by the model and led to the higher number of detected event boundaries. Nevertheless, correlation scores of the computational model for the noisy video segmentation were still better than the control models (see Supplementary Material). 

\subsection{Our results conform to the conceptualization of uncertainty framework}

The event segmentation ability of our model can be explained within the context of the uncertainty framework \citep{dayan2003uncertainty, dayan2006phasic}. According to this framework, a mental model can encounter two types of uncertainties: uncertainties because of the known unreliability of the predictive relationships and uncertainties above and beyond them, requiring a model change \citep{dayan2003uncertainty, dayan2006phasic, payzan2013neural, zhao2019pupil, angela2005uncertainty}. Expected uncertainty is related to the inherent stochasticity of the world that is expected from the perspective of the model. If an uncertainty is expected, one does not have to reset its internal models since they work for the current context. When a change is not explainable by the model, on the other hand, this is an unexpected uncertainty as the current model is not able to represent the current context and thus, the internal models are better change. In our proposed model, the threshold values of each event model determine their degrees of uncertainty or their representation powers for the current context. Prediction errors below this value are considered to be explainable within the dynamics of the model and therefore those changes form the expected uncertainty. If the observed changes are beyond the prediction capacity of the current model, in other words the current prediction error exceeds its threshold, those changes form the unexpected uncertainty and trigger a model reset. It has been observed that the unexpected uncertainties suppress the top-down expectations and increase the importance of newly acquired information \citep{dayan2003uncertainty, dayan2006phasic, farashahi2017metaplasticity, soltani2019adaptive, angela2005uncertainty}. Our model works similarly. When it observes an unexpected uncertainty, it suppresses the current event model and enters into a search period for learning a new sequence. 

\subsection{Event representations reveal perception of events}

We developed and employed a novel validation standard for our model based on event representations. For this aim, we conducted an online pairwise similarity judgment experiment and compared the similarity judgments of the participants for the segmented events with those of the computational model. We demonstrated that people were very reliable to the group decision in their similarity judgments, even when the events were displayed as PLDs \citep{Johansson1973VisualPerception}. In particular, mean correlation scores of each participant with all participants were 0.90 and 0.92 for the fine- and coarse-grained events, respectively. When we extracted the representations of the participants via various representation discovery techniques we developed, namely EER-CM/t-SNE, our model also received considerable correlation scores with the similarity judgments of the participants (0.435 and 0.614 for fine and coarse, respectively). Moreover, representations discovered by the EER-CM/t-SNE outperformed all the baseline models, directly using the information coming from the ground-truth data rather than the activations of a neural network. We believe that this finding is in line with the studies comparing the representations of neural networks and humans, showing their similarities, and using neural networks as a tool for capturing mental representations in neuroscience \citep{khaligh2014deep, mur2013human, tripp2017similarities} and psychology \citep{lake2015deep, dubey2015makes, kubilius2016deep, Hebart2020Revealing}. As a necessary reminder, EER-CM/t-SNE is partially dependent on the event segmentation mechanism of our model. The technique uses activations of a neural network model which receives various segmented events and discovers semantic relationship between those instances. This simple framework, namely segmenting event instances by monitoring the prediction errors and relating segments via an hierarchical mechanism implies the presence of a hierarchical (possibly recurrent) system in the event cognition. In fact, studies suggest that infants can find higher-level units based on the structural relationship between lower-level units \citep{levine2019finding, aslin2017statistical}. 

Despite its success in capturing the human event similarity judgments, EER-CM/t-SNE and EER-CM/PCA have an importation limitation. Basically, those models map each timestep to an event instance and, by this way, form the representation of each timestep as the training unfolds. Representations of each timesteps (frame-based representation) of a particular event are averaged out to extract a representation for that event (event-based representation). Although this technique is simple and understandable, the mean operation, namely averaging out, assigns the same credit for all timesteps in an event. However, this may not be the most suitable operation as humans remember (i.e., give an information processing advantage) the first and the end of a sequence better than its midpoint, known as the serial position effect \citep{Jonides2008Mind, Murdock1962Serial}. Further research might consider developing a method that aggregates frame-based representations based on the saliencies of the individual frames.

\subsection{Limitations}

Our model has a number of limitations. First of all, event models in our framework are distinct from one another, even though human actions are groupable. With the aim of addressing this issue, we connected different event segments by several representation discovery techniques. Further research can investigate how the relationship between the event instances can be utilized for achieving more robust event segmentation. Secondly, although our model can segment events in different lengths, it is not able to capture the part-whole relationship and use this information for the hierarchical segmentation. Even though we tried to exploit the event similarity judgments to capture the hierarchical organization of events, it was not successful. Recent research shows the possibility that hierarchical segmentation requires tracking higher-order statistical regularities between the event units \citep{Franklin2020Structured, Schapiro2013Neural}. Subsequent research can extend the proposed model in this way by learning temporal regularities between the event labels by a hierarchical recurrent neural network model. Further research can exploit the latent representation space proposed in this study for building a top-down signal, regulating behaviors of the event models. Such a latent representation space can be exploited to capture the event relationships, learn the new events faster, and segment events into parts more reliably.

At the first step of modeling event segmentation, we represented events by PLDs, which are utilized in event segmentation literature \citep{Reynolds2007Computational, Metcalf2017Modelling}. Further research can use naturalistic videos -rather than PLDs- to capture event segmentation abilities in certain event types, including human-object, human-human interactions, and scene-related changes. Since the current model trains all event models to find the next event model in the search period, using raw naturalistic videos may increase the computational load. Nevertheless, the current model can be scaled by an already trained object identification model such as AlexNet \citep{krizhevsky2012imagenet} or a dimensionality reduction algorithm such as PCA or autoencoders to reduce data dimensionality to achieve human-level event segmentation performance in the naturalistic videos. In the case of using sequences of images (i.e., videos) for representing events, convolutional neural networks (CNNs) might be more advantageous than the multi-layer perceptrons \citep{Goodfellow2016Deep, Haushofer2008Multivariate, Kriegeskorte2012Inverse}.

Despite these limitations, our model allows researchers to integrate various systems in a simple manner for exploring the prediction-error-based event segmentation and other perceptual and cognitive processes. This advantage is due to the semi-mechanistic nature of the model. For example, one can enrich the current model by defining the adaptive learning rates by comparing the thresholds of models and received prediction errors, and therefore, explore the effects of expected and unexpected uncertainties on learning \citep{dayan2003uncertainty, dayan2006phasic, farashahi2017metaplasticity, soltani2019adaptive, angela2005uncertainty}. Another possibility is to target prediction error signals and investigate their role in certain perceptual and cognitive processes. For instance, one can extend our model into a time perception model, which aggregates prediction error signals for estimating the duration of an event \citep{Basgol2021TimePerception, Fountas2020PredictiveProcessing}.

\subsection{Conclusion}

In this study, we developed a computational model of event segmentation, learning, and representation. Our model represents events by multi-layer perceptrons and manages them by monitoring their prediction error signals. By this way, our model can produce multimodal event segments in varying hierarchies via passive observation. We compared the capabilities of our model by event segmentation and representation experiments.

\begin{itemize}
\item We evaluated fine-grained and coarse-grained event segmentation performances of the model with two videos, representing various human behaviors by PLDs \citep{Johansson1973VisualPerception}. We then compared its performance by online psychological experiments. We showed that the proposed model received 0.254 and 0.196 point-biserial correlation scores for the fine- and coarse-grained segmentation, respectively.

\item With the aim of examining whether the event segments that are produced by our model are interpretable and extractable from the activations of neural networks, we proposed a new validation technique inspired from the literature \citep{Peterson2018Evaluating}. We approximated the event representations of our model by several techniques of participants by an event similarity judgment experiment and extracted their event representation spaces. We compared the event representation spaces of our model with those of the participants and received the correlation scores of 0.435 and 0.614 for the fine- and coarse-grained segments, respectively.

\item Our results suggest that a model based on monitoring the prediction error signals can capture the event segmentation decisions of humans, and multi-layer perceptrons can capture the internal representation space.

\item From a broader perspective, our results confirm an already held theory, event segmentation theory \citep{Zacks2007EventSegmentation}, and computational and robotic models in the literature \citep{Gumbsch2016Learning, Gumbsch2017ComputationalModelDyn, Gumbsch2019Autonomous, Reynolds2007Computational, Metcalf2017Modelling}. Moreover, they imply a hierarchical mechanism relating event segments \citep{levine2019finding, aslin2017statistical} and present the possibility of tracking higher-order statistical regularities between event segments in a more opaque way \citep{Franklin2020Structured, Schapiro2013Neural}. The present model shows how perceptual and cognitive abilities can be modeled with maintaining interpretability and without the need of human-crafted labels.

\end{itemize} 

\section{Acknowledgement}

This work was supported in part by the Bogazici Research Projects (BAP), Project titled “A Computational Model of Event Learning and Segmentation: Event Granularity, Sensory Reliability and Expectation," under Grant 20A01P2.

\bibliography{BibliographyFile}

\end{document}